\documentclass[12pt]{article}
\pdfoutput=1

\usepackage[utf8]{inputenc}
\usepackage[left=2.55cm, right=2.55cm, top=2.55cm, bottom=2.55cm]{geometry}
\usepackage{amsmath,amssymb,amsbsy}
\usepackage{slashed}
\usepackage{xcolor}
\usepackage{graphicx}
\usepackage{url}
\usepackage{cancel}
\usepackage{cite}
\usepackage[colorlinks=true,allcolors=darkpurple,pdfborder={0 0 0},linktocpage=false]{hyperref}
\usepackage{tabularx,booktabs}
\usepackage{multicol}
\usepackage{units}
\usepackage{xspace}
\usepackage[labelfont=bf]{caption}
\usepackage[section]{placeins}

\definecolor{darkred}{rgb}{0.6,0,0}
\definecolor{darkpurple}{rgb}{0.5,0,0.5}
\def\vev#1{\left\langle #1\right\rangle}
\def\hc{\text{h.c.}}

\def\Tr{\text{Tr}}

\newcommand{\AddrIFIC}{%
  Instituto de F\'{i}sica Corpuscular, CSIC-Universitat de Val\`{e}ncia, 46980 Paterna, Spain}

\newcommand{\AddrFISTEO}{%
  Departament de F\'{\i}sica Te\`{o}rica, Universitat de Val\`{e}ncia, 46100 Burjassot, Spain}

\begin{document}

\vspace*{-2cm}
\begin{flushright}
IFIC/21-27 \\
\vspace*{2mm}
%\today
\end{flushright}

\begin{center}
\vspace*{15mm}

\vspace{1cm}
{\Large \bf 
An ultraviolet completion for the Scotogenic model
} \\
\vspace{1cm}

{\bf Pablo Escribano$^{\text{a}}$, Avelino Vicente$^{\text{a,b}}$}

 \vspace*{.5cm} 
 $^{(\text{a})}$ \AddrIFIC \\\vspace*{.2cm} 
 $^{(\text{b})}$ \AddrFISTEO

 \vspace*{.3cm} 
\href{mailto:pablo.escribano@ific.uv.es}{pablo.escribano@ific.uv.es}, \href{mailto:avelino.vicente@ific.uv.es}{avelino.vicente@ific.uv.es}
\end{center}

\vspace*{10mm}
\begin{abstract}\noindent\normalsize
  The Scotogenic model is an economical scenario that generates
  neutrino masses at the 1-loop level and includes a dark matter
  candidate. This is achieved by means of an \emph{ad-hoc}
  $\mathbb{Z}_2$ symmetry, which forbids the tree-level generation of
  neutrino masses and stabilizes the lightest $\mathbb{Z}_2$-odd
  state. Neutrino masses are also suppressed by a quartic coupling,
  usually denoted by $\lambda_5$. While the smallness of this
  parameter is natural, it is not explained in the context of the
  Scotogenic model. We construct an ultraviolet completion of the
  Scotogenic model that provides a natural explanation for the
  smallness of the $\lambda_5$ parameter and induces the
  $\mathbb{Z}_2$ parity as the low-energy remnant of a global $\rm
  U(1)$ symmetry at high energies. The low-energy spectrum contains,
  besides the usual Scotogenic states, a massive scalar and a massless
  Goldstone boson, hence leading to novel phenomenological predictions
  in flavor observables, dark matter physics and colliders.
\end{abstract}

\section{Introduction}
\label{sec:intro}

The smallness of neutrino masses can be understood if they are
radiatively
generated~\cite{Zee:1980ai,Cheng:1980qt,Zee:1985id,Babu:1988ki}. Indeed,
if neutrinos are massless at tree-level but become massive at higher
loop orders, a natural suppression for their masses emerges. Many
radiative neutrino mass models exist~\cite{Cai:2017jrq}. They extend
the Standard Model (SM) particle content with new states and, very
often, also with new symmetries that prevent neutrinos from becoming
massive at tree-level. In general, the phenomenology of these models
is very rich, due to the presence of new states with sizable couplings
to the SM particles.

The Scotogenic model~\cite{Ma:2006km} is arguably one of the most
popular radiative neutrino mass models. In this economical setup, the
SM particle content is extended with new \emph{inert} scalar doublets,
that couple only to leptons and do not acquire vacuum expectation
values (VEVs), and fermion singlets. In the usual version of the
model, one inert doublet and three fermion singlets are introduced,
although other choices are
possible~\cite{Hehn:2012kz,Fuentes-Martin:2019bue} and more general
scenarios can be considered~\cite{Escribano:2020iqq}. The new states
are odd under an additional $\mathbb{Z}_2$ symmetry, under which the
SM states are assumed to be even. This symmetry has a twofold purpose:
\begin{itemize}
  \item It forbids the tree-level generation of neutrino masses, which
    are nevertheless generated at the 1-loop level. These are
    naturally small, not only due to the usual loop suppression, but
    also because they are proportional to a small parameter in the
    scalar potential, the so-called $\lambda_5$ quartic coupling. The
    presence of this parameter is required to break lepton number in
    two units, and therefore its smallness is natural in the sense of
    't Hooft~\cite{tHooft:1979rat}, although it is not explained in
    the context of the Scotogenic model.
  \item The lightest $\mathbb{Z}_2$-odd state is stable and can thus
    be a valid dark matter (DM) candidate. This role can be played by
    a scalar state or by the lightest fermion singlet.
\end{itemize}
In this letter we consider an ultraviolet completion of the Scotogenic
model that provides a natural explanation for the smallness of the
$\lambda_5$ parameter. In addition, the \emph{dark} $\mathbb{Z}_2$
parity present in the model emerges at low energies from the breaking
of a global $\rm U(1)$ lepton number symmetry present at high
energies~\footnote{The generation of the $\mathbb{Z}_2$ Scotogenic
parity from the breaking of a $\rm U(1)$ lepton number symmetry has
been discussed in~\cite{Ma:2015xla,CentellesChulia:2019gic}.}. As a
result of this, the low-energy theory will consist of the Scotogenic
model supplemented with two additional scalar states: a massive scalar
and a massless Goldstone boson, the
\emph{majoron}~\cite{Chikashige:1980ui,Gelmini:1980re,Schechter:1981cv,Aulakh:1982yn}. Therefore,
our scenario explains some of the open questions of the original
Scotogenic model and leads to novel phenomenological consequences due
to the presence of these states, as will be shown below.

The rest of the manuscript is organized as
follows. Section~\ref{sec:UV} presents the complete ultraviolet
theory, whereas Section~\ref{sec:IR} derives the resulting effective
theory at the electroweak scale, as well as its most relevant
features. Section~\ref{sec:pheno} discusses some phenomenological
consequences of our construction, mostly focusing on the role of the
majoron. Finally, we summarize our work in
Section~\ref{sec:conclusions}.

\section{Ultraviolet theory}
\label{sec:UV}

The particle content of the Scotogenic model~\cite{Ma:2006km}
includes, besides the usual Standard Model (SM) fields, three
generations of fermions $N$, transforming as $\left(
\mathbf{1},0\right)$ under $\rm \left( \rm SU(2)_L , U(1)_Y \right)$,
and one scalar $\eta$, transforming as $\left(
\mathbf{2},1/2\right)$. Therefore, the model contains two scalar
doblets, the usual Higgs doublet $H$ and the new doublet $\eta$,
decomposed in terms of their $\rm SU(2)_L$ components as
\begin{equation}
  H = \begin{pmatrix}
    H^+ \\
    H^0 
  \end{pmatrix} \, , \quad \eta = \begin{pmatrix}
    \eta^+ \\
    \eta^0 
  \end{pmatrix} \, .
\end{equation}
Our ultraviolet enlarges the particle content of the Scotogenic model
with two new multiplets: a scalar $\rm SU(2)_L$ triplet $\Delta$,
written as a $2 \times 2$ matrix as
\begin{equation}
  \Delta = 
  \begin{pmatrix}
    \Delta^+ / \sqrt{2} & \Delta^{++} \\
    \Delta^0 & - \Delta^+ /\sqrt{2}
  \end{pmatrix} \, ,
\end{equation}
and a scalar singlet $S$. In addition, instead of the usual
$\mathbb{Z}_{2}$ Scotogenic parity, a global $\rm U(1)_L$ symmetry,
where $L$ stands for lepton number, is introduced. A similar model
with a gauge $\rm U(1)_L$ symmetry can be built, but we leave this
version of our setup for future work.  Table~\ref{tab:ParticleContent}
shows the scalar and leptonic fields of the model and their
representations under the gauge and global symmetries.

{
\renewcommand{\arraystretch}{1.4}
\begin{table}[tb]
  \centering
  \begin{tabular}{|c|c||ccc|c|}
      \hline
      \textbf{Field} & \textbf{Generations} & \textbf{$\rm SU(3)_c$} & \textbf{$\rm SU(2)_L$} & \textbf{$\rm U(1)_Y$} & \textbf{$\rm U(1)_L$} \\
      \hline
      $\ell_L$      &  3  &  1  &  2  &  -1/2  &  $1$   \\
      $e_R$    &  3  &  1  &  1  &  -1    &  $1$   \\
      $N$      &  3  &  1  &  1  &  0     &  $\frac{1}{2}$  \\
      \hline
      $H$      &  1  &  1  &  2  &  1/2   &  $0$   \\
      $\eta$   &  1  &  1  &  2  &  1/2   &  $-\frac{1}{2}$  \\
      $\Delta$ &  1  &  1  &  3  &  1     &  $-1$  \\  
      $S$      &  1  &  1  &  1  &  0     &  $1$   \\
      \hline
  \end{tabular}
  \caption{\label{tab:ParticleContent}
    Lepton and scalar particle content of the model and their representations under the gauge and global symmetries. $\ell_L$ and $e_R$ are the SM left- and right-handed leptons, respectively, and $H$ is the SM Higgs doublet.}
\end{table}
}

The complete Lagrangian of the theory can be written as
\begin{equation}
  \mathcal{L} = \mathcal{L}_{\rm SM} + y \, \overline{N} \, \eta \, i \sigma_2 \, \ell_L + \kappa \, S^* \, \overline{N^c} N + \hc - \mathcal{V}_{\rm UV} \, ,
\end{equation}
where $\mathcal{L}_{\rm SM}$ is the SM Lagrangian (without the scalar
potential), the second and third terms are Yukawa terms and
$\mathcal{V}_{\rm UV}$ is the scalar potential, given by
\begin{equation}
  \begin{split}
    \mathcal{V}_{\rm UV} & = m_H^2 H^\dagger H + m_S^2 S^* S + m_\eta^2 \eta^\dagger \eta + m_\Delta^2 \Tr \left( \Delta^\dagger \Delta \right) + \frac{1}{2} \lambda_1 \left( H^\dagger H \right)^2 + \frac{1}{2} \lambda_S \left( S^* S \right)^2 + \frac{1}{2} \lambda_2 \left( \eta^\dagger \eta \right)^2 \\
    & + \frac{1}{2} \lambda_{\Delta 1} \Tr \left( \Delta^\dagger \Delta \right)^2 + \frac{1}{2} \lambda_{\Delta 2} \left( \Tr \, \Delta^\dagger \Delta \right)^2 + \lambda_3^S \left( H^\dagger H \right) \left( S^* S \right) + \lambda_3 \left( H^\dagger H \right) \left( \eta^\dagger \eta \right) \\
    & + \lambda_3^\Delta \left( H^\dagger H \right) \Tr \left( \Delta^\dagger \Delta \right) + \lambda_3^{\eta S} \left( \eta^\dagger \eta \right) \left( S^* S \right) + \lambda_3^{\eta \Delta} \left( \eta^\dagger \eta \right) \Tr \left( \Delta^\dagger \Delta \right) + \lambda_3^{S \Delta} \left( S^* S \right) \Tr \left( \Delta^\dagger \Delta \right) \\
    & + \lambda_4 \left( H^\dagger \eta \right) \left( \eta^\dagger H \right) + \lambda_4^\Delta \left( H^\dagger \Delta^\dagger \Delta \, H \right) + \lambda_4^{\eta \Delta} \left( \eta^\dagger \Delta^\dagger \Delta \, \eta \right) \\
    & + \left[ \lambda_{H S \Delta} S \left( H^\dagger \Delta \, i \sigma_2 H^* \right) + \mu \left( \eta^\dagger \Delta \, i \sigma_2 \, \eta^* \right) + \hc \right] \, . \label{eq:full-potential}    
  \end{split}
\end{equation}
Notice that the Lagrangian terms $\left(H^\dagger \, \eta \right)^2$,
$H^\dagger \Delta \eta^\dagger$, $H^\dagger \Delta \eta^\dagger S$ and
$H^\dagger \eta S$ are allowed by the gauge symmetries, but not by
lepton number.

\section{Low-energy theory}
\label{sec:IR}

In the following we assume that $m_\Delta$ is much larger than any
other mass scale in the model. We can therefore determine the
effective Lagrangian of the theory at energies much below $m_\Delta$
by integrating out the heavy $\Delta$ triplet and expanding the result
in powers of $1 / m_\Delta$. Since we are only interested in working
at tree-level, this can be easily achieved by following the method
described in~\cite{deBlas:2014mba}. In the following we assume that,
before electroweak symmetry breaking, the scalar mass matrix has only
one negative eigenvalue, $-\mu_H^2$. Its associated eigenvector is a
$\left( \mathbf{2},1/2 \right)$ scalar field $H$, which we identify
with the SM Higgs doublet. We assume that the dimensionful parameter
$\mu$ that appears with the dimension-three operator in the scalar
potential is at most of the size of the mass of the triplet,
$m_\Delta$. These assumptions lead to a decoupling scenario and allow
us to perform the integration in the electroweak symmetric phase,
which is extremely convenient.

After integrating out the $\Delta$ triplet, the scalar potential
suffers several changes. In particular, the scalar potencial of the
low-energy theory reads as follows:
\begin{align}
    \mathcal{V}_{\rm IR} & = m_H^2 \, H^\dagger H + m_S^2 \, S^* S + m_\eta^2 \, \eta^\dagger \eta + \left( H^\dagger H \right)^2 \left[\frac{\lambda_1}{2} - \frac{\left| \lambda_{HS\Delta}\right|^2}{m_\Delta^2} \left( S^* S \right) \right] + \frac{\lambda_S}{2} \left( S^* S \right)^2 \nonumber \\
    & + \left( \eta^\dagger \eta \right)^2 \left(\frac{\lambda_2}{2} - \frac{\left| \mu\right|^2}{m_\Delta^2} \right) + \lambda_3^S \left( H^\dagger H \right) \left( S^* S \right) + \lambda_3 \left( H^\dagger H \right) \left( \eta^\dagger \eta \right) + \lambda_3^{\eta S} \left( \eta^\dagger \eta \right) \left( S^* S \right) \nonumber \\
    & + \lambda_4 \left( H^\dagger \eta \right) \left( \eta^\dagger H \right) - \left[ \frac{\lambda_{HS\Delta} \, \mu^*}{m_\Delta^2} S \left( H^\dagger \eta \right)^2 + \hc \right] + \mathcal{O}\left(\frac{1}{m_{\Delta}^{4}}\right) \, .
  \label{eq:EFT-potential}
\end{align}
We now write the Lagrangian in the broken phase. First, we decompose
the neutral $H^0$ and $S$ fields as
\begin{equation}
  H^0 = \frac{1}{\sqrt{2}} \left( v_H + \phi + i \, A \right) \, ,
  \quad S = \frac{1}{\sqrt{2}} \left( v_S + \rho + i \, J \right) \, .
  \label{eq:neutral_fields}
\end{equation}
This defines the vacuum expectation values (VEVs) of $H^0$ and $S$,
$\frac{v_H}{\sqrt{2}}$ and $\frac{v_S}{\sqrt{2}}$, respectively. The
tadpole equations of the potential evaluated in these VEVs are
\begin{align}
    & \left. \frac{d\mathcal{V}_{\rm IR}}{dH^0}\right|_{\vev{H^0} = \frac{v_H}{\sqrt{2}}, \vev{S} = \frac{v_S}{\sqrt{2}}} = \frac{v_H^*}{\sqrt{2}} \left( m_H^2 + \lambda_1 \frac{\left| v_H \right|^2}{2} + \lambda_3^S \frac{\left| v_S \right|^2}{2} - \frac{\left| v_H \right|^2 \left| v_S \right|^2 \left| \lambda_{H S \Delta} \right|^2}{2 m_\Delta^2} \right) = 0 \, , \label{eq:tad1} \\
    & \left. \frac{d\mathcal{V}_{\rm IR}}{dS}\right|_{\vev{S} = \vev{H^0} = \frac{v_H}{\sqrt{2}}, \frac{v_S}{\sqrt{2}}} = \frac{v_S^*}{\sqrt{2}} \left( m_S^2 + \lambda_3^S \frac{\left| v_H \right|^2}{2} + \lambda_S \frac{\left| v_S \right|^2}{2} - \frac{\left| v_H \right|^4 \left| \lambda_{H S \Delta} \right|^2}{4 m_\Delta^2} \right) = 0 \, . \label{eq:tad2}
\end{align}
One can easily find analytical solutions for the VEVs, expanding them
in powers of $m_\Delta^{-2}$, $v^2_{H, S} = v^{2 \, (0)}_{H, S} + v^{2
  \, (1)}_{H, S} + \dots$, where $v^{2 \, (n)}_{H, S}$ is of order
$\left(1/m_\Delta\right)^{2n}$. Two comments are now in order. First,
the $S$ VEV breaks the global $\rm U(1)_L$ symmetry, but leaves a
remnant $\mathbb{Z}_2$ symmetry:
\begin{equation*}
{\rm U(1)_L} \, \xrightarrow{\quad v_S \quad} \, \mathbb{Z}_2
\end{equation*}
All the fields in the model are even under the remnant $\mathbb{Z}_2$,
with the only exception of $N$ and $\eta$, which are odd. This is
precisely the usual dark parity in the Scotogenic model, obtained in
our setup as a residual symmetry after the breaking of lepton
number. We also notice that the $S \left( H^\dagger \eta \right)^2$
term in the scalar potential of Eq.~\eqref{eq:EFT-potential} has
exactly the form of the $\lambda_5$ term of the Scotogenic model once
the singlet $S$ acquires a VEV. So, given that the mass of the triplet
is much larger than any other dimensionful parameter in the model, we
have a natural explanation of the smallness of the effective
$\lambda_5$ coupling,
\begin{equation} \label{eq:lam5}
  \frac{\lambda_5}{2} \equiv - \frac{\lambda_{H S \Delta} \, \mu^* \, v_S}{\sqrt{2} \, m_\Delta^2} \ll 1 \, .
\end{equation}

We turn now our attention to the scalar mass spectrum of the
model. Let us first consider the $\mathbb{Z}_2$-even scalars. Assuming
that CP is conserved in the scalar sector, the CP-even states $\phi$
and $\rho$ do not mix with the CP-odd states $A$ and $J$. In the bases
$\{\phi,\rho\}$ and $\{A,J\}$, their mass matrices read
\begin{equation}
  \mathcal{M}^2_R = \begin{pmatrix}
                       m_H^2 + \frac{3 \, v_{H}^2}{2} \, \lambda_1 + \frac{v_S^2}{2} \, \lambda_3^S - \frac{3 \, v_H^2 \, v_S^2 \left|\lambda_{H S \Delta}\right|^{2}}{2 \, m_{\Delta}^{2}} & v_H \, v_S \left( \lambda_3^S - \frac{v_H^2 \left|\lambda_{H S \Delta}\right|^{2}}{m_{\Delta}^{2}} \right) \\
                       v_H \, v_S \left( \lambda_3^S - \frac{v_H^2 \left|\lambda_{H S \Delta}\right|^{2}}{m_{\Delta}^{2}} \right) & m_S^2 + \frac{3 \, v_S^2}{2} \, \lambda_S + \frac{v_H^2}{2} \, \lambda_3^S - \frac{v_H^4 \left|\lambda_{H S \Delta}\right|^{2}}{4 \, m_{\Delta}^{2}}
                     \end{pmatrix} \, ,
\end{equation}
and
\begin{equation}
  \mathcal{M}^2_I = \begin{pmatrix}
                       m_H^2 + \frac{v_{H}^2}{2} \, \lambda_1 + \frac{v_S^2}{2} \, \lambda_3^S - \frac{v_H^2 \, v_S^2 \left|\lambda_{H S \Delta}\right|^{2}}{2 \, m_{\Delta}^{2}}  & 0 \\
                       0 & m_S^2 + \frac{v_S^2}{2} \, \lambda_S + \frac{v_H^2}{2} \, \lambda_3^S - \frac{v_H^4 \left|\lambda_{H S \Delta}\right|^{2}}{4 \, m_{\Delta}^{2}}
                     \end{pmatrix} \, ,
\end{equation}
respectively. Using now the tadpole equations in Eqs.~\eqref{eq:tad1}
and \eqref{eq:tad2} one can simplify these matrices notably. In fact,
the CP-odd matrix $\mathcal{M}^2_I$ is exactly zero once the
minimization equations are used. This is not surprising. One of the
states ($A$) is the would-be Goldstone that becomes the longitudinal
component of the $Z$ boson and makes it massive, while the other ($J$)
is the majoron, a (physical) massless Goldstone boson associated to
the spontaneous breaking of lepton
number~\cite{Chikashige:1980ui,Gelmini:1980re,Schechter:1981cv,Aulakh:1982yn}. The
CP-even matrix $\mathcal{M}^2_R$ becomes
\begin{equation}
  \mathcal{M}^2_R = \begin{pmatrix}
                       v_H^2 \left( \lambda_1 - \frac{v_S^2 \left| \lambda_{H S \Delta} \right|^2}{m_\Delta^2} \right) & v_H \, v_S \left( \lambda_3^S - \frac{v_H^2 \left| \lambda_{H S \Delta} \right|^2}{m_\Delta^2} \right) \\
                       v_H \, v_S \left( \lambda_3^S - \frac{v_H^2 \left| \lambda_{H S \Delta} \right|^2}{m_\Delta^2} \right) & v_S^2 \, \lambda_S
                     \end{pmatrix} \, .
\end{equation}
This matrix can be brought to diagonal form as $V_R^T \,
\mathcal{M}^2_R \, V_R = \text{diag} \left( m_h^2 , m_\Phi^2 \right)$,
with
\begin{equation} \label{eq:mixing}
  V_R = \begin{pmatrix}
    \cos \alpha & - \sin \alpha \\
    \sin \alpha & \cos \alpha
  \end{pmatrix} \, , \quad \tan (2 \alpha) = \frac{2 \, \left(\mathcal{M}^2_R\right)_{12}}{\left(\mathcal{M}^2_R\right)_{11}-\left(\mathcal{M}^2_R\right)_{22}} \approx 2 \, \frac{\lambda_3^S}{\lambda_S} \, \frac{v_H}{v_S} \, ,
\end{equation}
where the mixing angle $\alpha$ is given at leading order in
$1/m_\Delta^2$ and the approximation assumes $v_H \ll v_S$. Therefore,
a mixing exists between the real parts of the neutral component of the
$H$ doublet, $\phi$, and of the $S$ singlet, $\rho$. The mixing is
however suppressed by the $v_H/v_S$ ratio which is assumed to be much
smaller than $1$. The lightest of the resulting two mass eigenstates
is to be identified with the Higgs-like state $h$, with mass $m_h
\approx 125$ GeV, discovered at the LHC. An additional CP-even state
is present in the spectrum, and we will denote it as $\Phi$. We
consider now the $\mathbb{Z}_2$-odd scalars. We decompose the neutral
$\eta$ component as
\begin{equation}
  \eta^0 = \frac{1}{\sqrt{2}} \left( \eta_R + i \, \eta_I \right) \, .
\end{equation}
Again, assuming the conservation of CP in the scalar sector, $\eta_R$
and $\eta_I$ do not mix. Their masses, as well as the mass of the
charged $\eta$ component, are given by
\begin{align}
  m^2_{\eta_R} =& \, m_\eta^2 + \lambda_3^S \, \frac{v_S^2}{2} + \left( \lambda_3 + \lambda_4 - \frac{2 \, \lambda_{H S \Delta} \, \mu \, v_{S}}{\sqrt{2} \, m_{\Delta}^{2}} \right) \frac{v_H^2}{2} \, , \\
  m^2_{\eta_I} =& \, m_\eta^2 + \lambda_3^S \, \frac{v_S^2}{2} + \left( \lambda_3 + \lambda_4 + \frac{2 \, \lambda_{H S \Delta} \, \mu \, v_{S}}{\sqrt{2} \, m_{\Delta}^{2}} \right) \frac{v_H^2}{2} \, , \\
  m^2_{\eta^+} =& \, m_\eta^2 + \lambda_3 \, \frac{v_H^2}{2} + \lambda_3^{\eta S} \, \frac{v_S^2}{2} \, .
\end{align}
As in the usual Scotogenic model, the mass difference between $\eta_R$
and $\eta_I$ is controlled by the effective $\lambda_5$ coupling,
\begin{equation}
  m^2_{\eta_R} - m^2_{\eta_I} = - \frac{4 \, \lambda_{H S \Delta} \, \mu \, v_{S}}{\sqrt{2} \, m_{\Delta}^{2}} \, \frac{v_H^2}{2} \equiv \lambda_5 \, v_H^2 \, .
\end{equation}

\begin{figure}[tb!]
  \centering
  \includegraphics[width=0.45\linewidth]{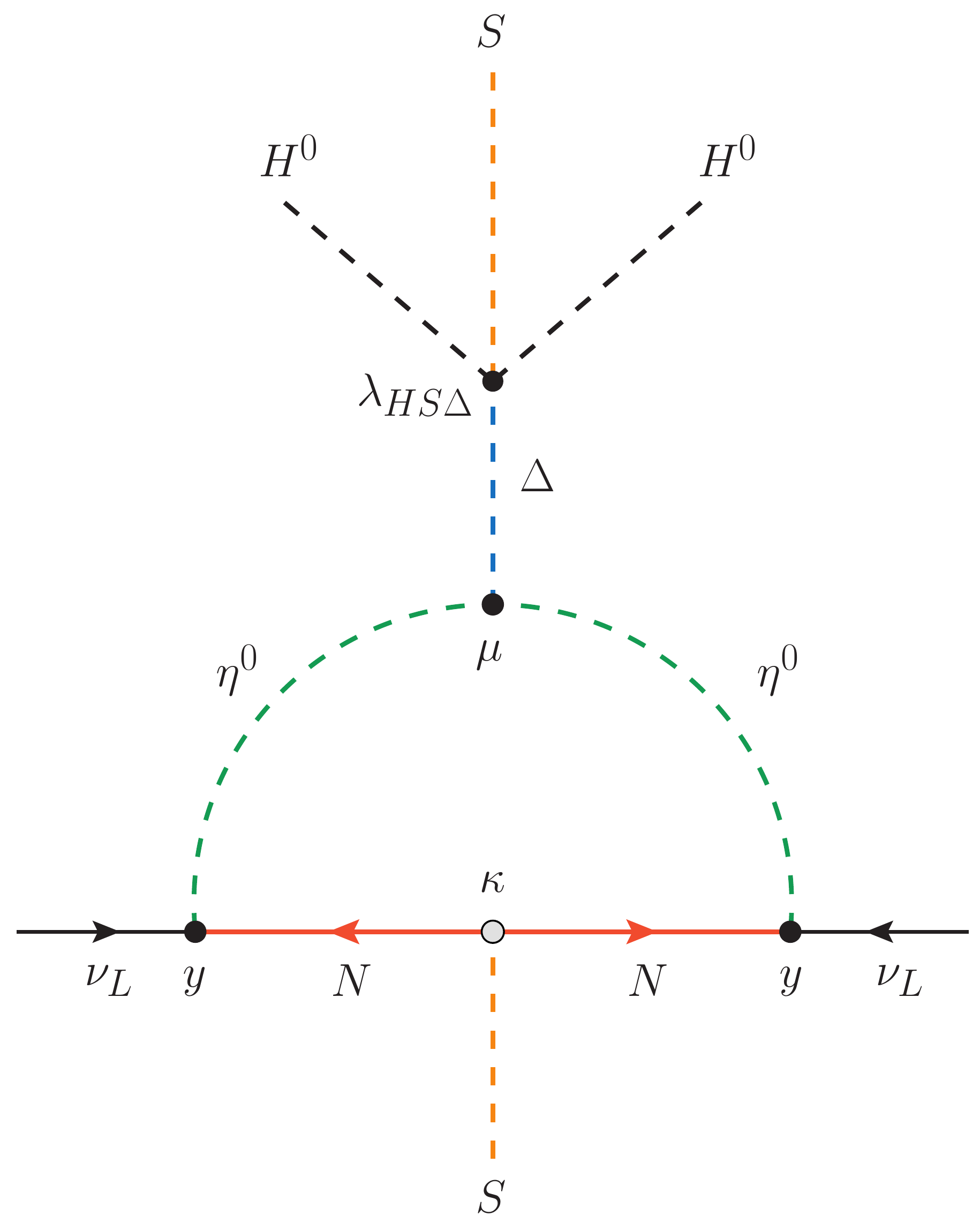}
  \includegraphics[width=0.45\linewidth]{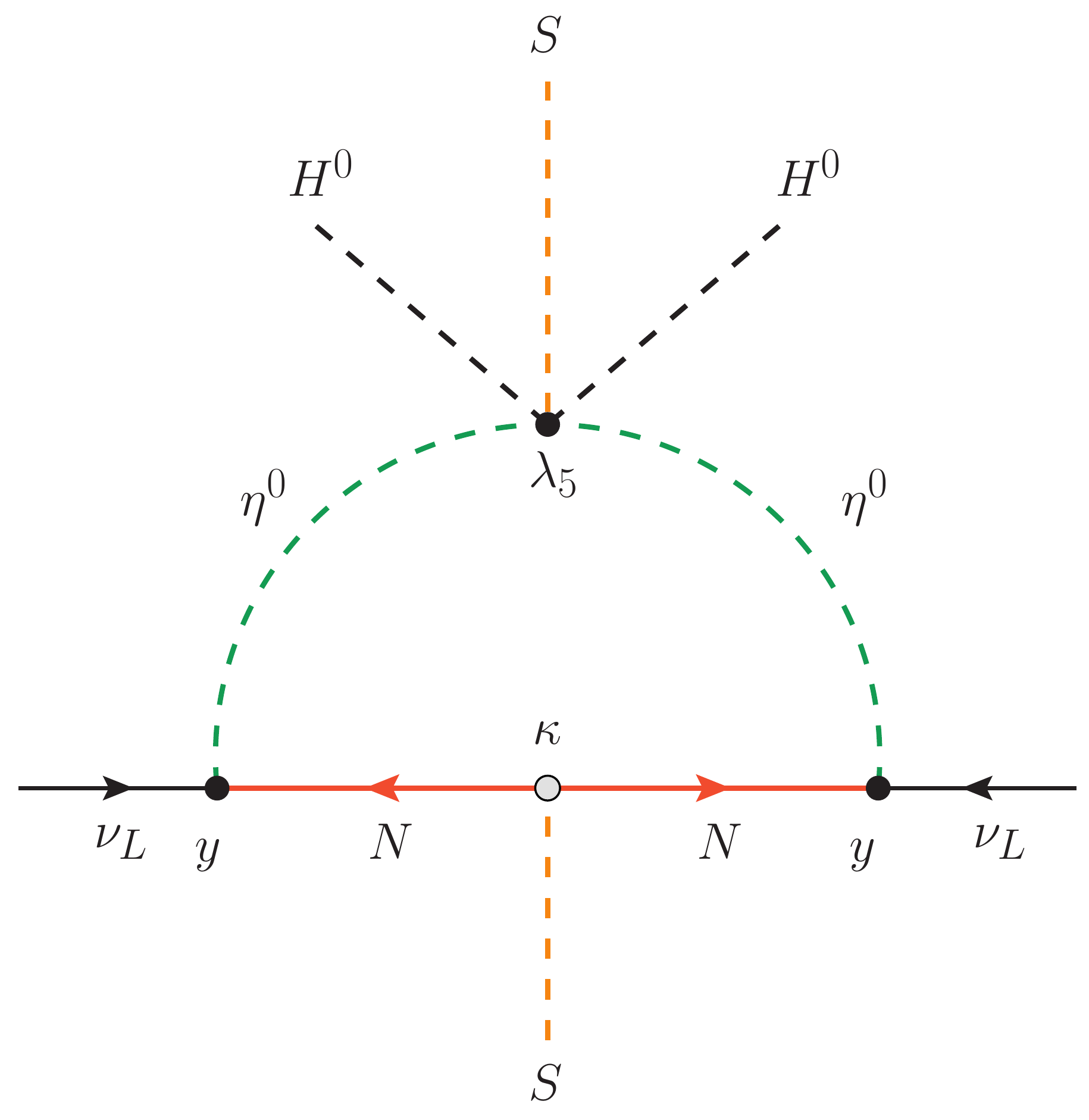}
  \caption{Neutrino mass generation in the ultraviolet (left) and
    low-energy theories (right). The effective $\lambda_5$ coupling
    enables the generation of neutrino masses at the 1-loop level in
    the low-energy theory.
    \label{fig:numass}
    }
\end{figure}

Finally, we comment on neutrino masses. The breaking of the $\rm
U(1)_L$ global symmetry generates a Majorana mass term for the $N$
fermions, $\frac{M_N}{2} \, \overline{N^c} \, N$, with
\begin{equation}
  M_N = \sqrt{2} \, \kappa \, v_S \, .
\end{equation}
In the following, we will work in the basis in which this matrix is
diagonal. Therefore, $M_{N_n}$ (with $n=1,2,3$) represent the physical
masses of the $N$ fermionic singlets and, as a consequence of this,
for singlet masses above the electroweak scale, one naturally has $v_S
\gg v_H$. After the electroweak and $\rm U(1)_L$ symmetries are
broken, non-zero Majorana neutrino masses are induced at the 1-loop
level, as shown in Fig.~\ref{fig:numass}. The left-hand side of this
figure displays the relevant diagram for the generation of neutrino
masses in the ultraviolet theory, while the right-hand side shows the
equivalent diagram at low energies, once $\Delta$ is integrated out
and an effective $\lambda_5$ coupling is obtained. The resulting
diagram is the usual Scotogenic loop and one obtains the well-known
expression for the neutrino mass matrix
\begin{equation} \label{eq:mnu31}
\left(m_{\nu}\right)_{\alpha \beta} = \frac{\lambda_{5} \, v_H^2}{32 \pi^{2}} \sum_{n} \frac{y_{n \alpha} \, y_{n \beta}}{M_{N_n}} \left[ \frac{M_{N_n}^{2}}{m_{0}^{2} - M_{N_n}^{2}} + \frac{M_{N_n}^{4}}{\left(m_{0}^{2} - M_{N_n}^{2}\right)^{2}} \log \frac{M_{N_n}^{2}}{m_{0}^{2}} \right] \, ,
\end{equation}
with $m_0^2 = m_\eta^2 + (\lambda_3 + \lambda_4) \, v_H^2 / 2$. This
expression agrees with~\cite{Ma:2006km} up to a factor of $1/2$ that
was missing in the original reference.

\section{Phenomenological consequences}
\label{sec:pheno}

In this Section we explore some of the phenomenological consequences
of our model and highlight some distinctive features, not present in
the \textit{minimal} Scotogenic model. In addition to the usual fields
present in the Scotogenic model, the particle spectrum of the
low-energy theory contains a massive CP-even scalar, $\Phi$, and a
massless Goldstone boson, the majoron $J$. The presence of this new
degree of freedom may have important phenomenological consequences.

\subsubsection*{Majoron couplings to charged leptons}

\begin{figure}[tb!]
  \centering
  \includegraphics[width=0.5\linewidth]{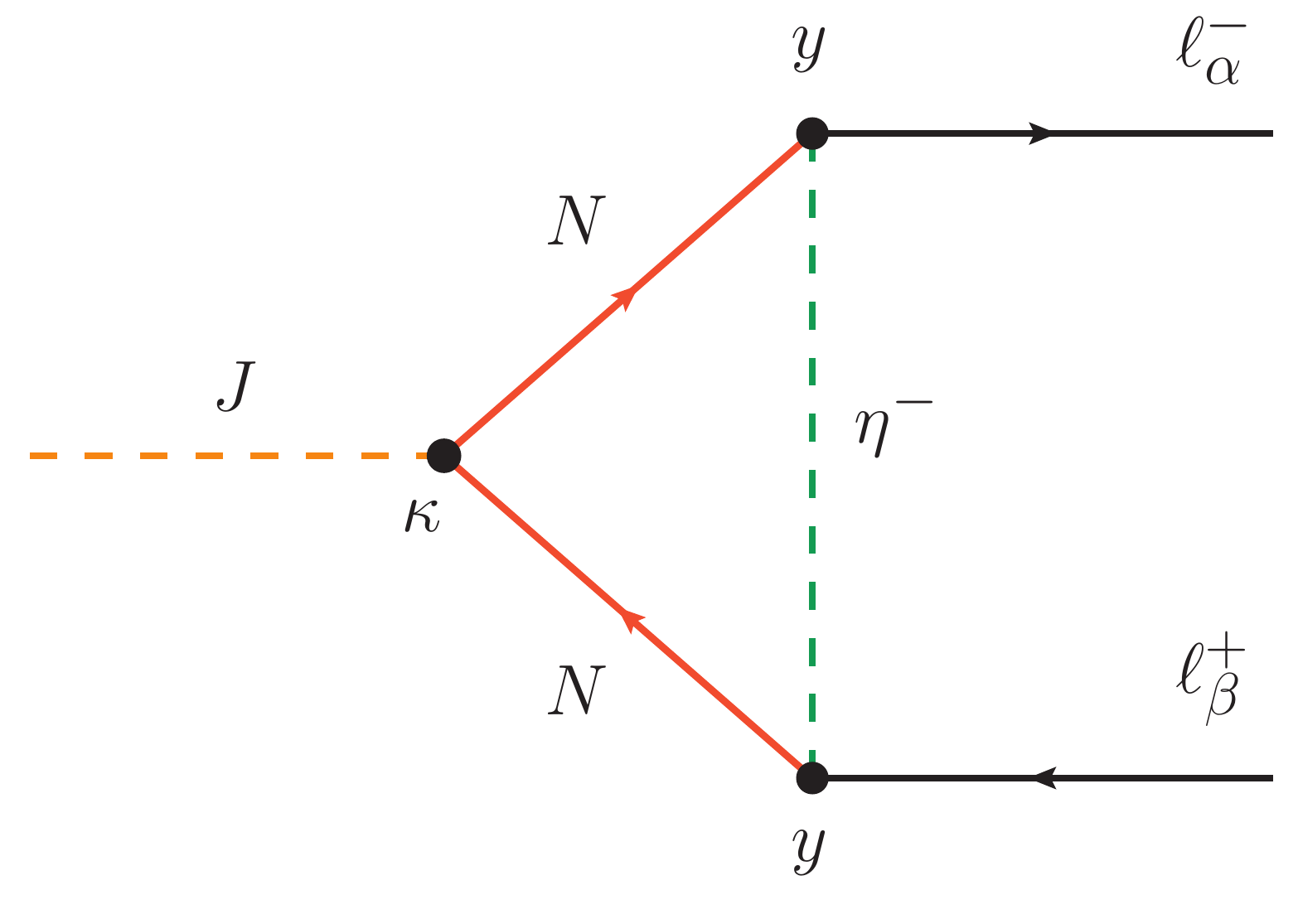}
  \caption{Loop induced majoron couplings to the SM charged leptons.
    \label{Maj-chlep:diag}
    }
\end{figure}

Majoron couplings to a pair of charged leptons are induced at the 1-loop
level by the Feynman diagram in Fig.~\ref{Maj-chlep:diag}. Neglecting
corrections proportional to the charged lepton massess, the resulting
interaction vertex is given by
\begin{equation}
  \mathcal{L}_{J \ell \ell} = - \frac{i \, J}{16 \, \pi^2 \, v_S} \, \overline{\ell} \left( M_\ell \, y^\dagger \, \Gamma \, y \, P_L - y^\dagger \, \Gamma \, y \, M_\ell \, P_R  \right) \ell \, , 
\end{equation}
where $M_\ell = \text{diag}(m_e,m_\mu,m_\tau)$ and we have defined
\begin{equation}
  \Gamma_{m n} = \frac{M_{N_n}^2}{\left( M_{N_n}^2 - m^2_{\eta^+} \right)^2} \left( M_{N_n}^2 - m^2_{\eta^+} + m^2_{\eta^+} \, \log \frac{m^2_{\eta^+}}{M_{N_n}^2} \right) \delta_{mn}.
\end{equation}
These results are analogous to those found in the type-I seesaw with
spontaneous violation of lepton number, in which majoron couplings to
charged leptons are also induced at the 1-loop level, see for
instance~\cite{Heeck:2019guh}.

\begin{figure}[tb!]
  \centering
  \includegraphics[width=0.6\linewidth]{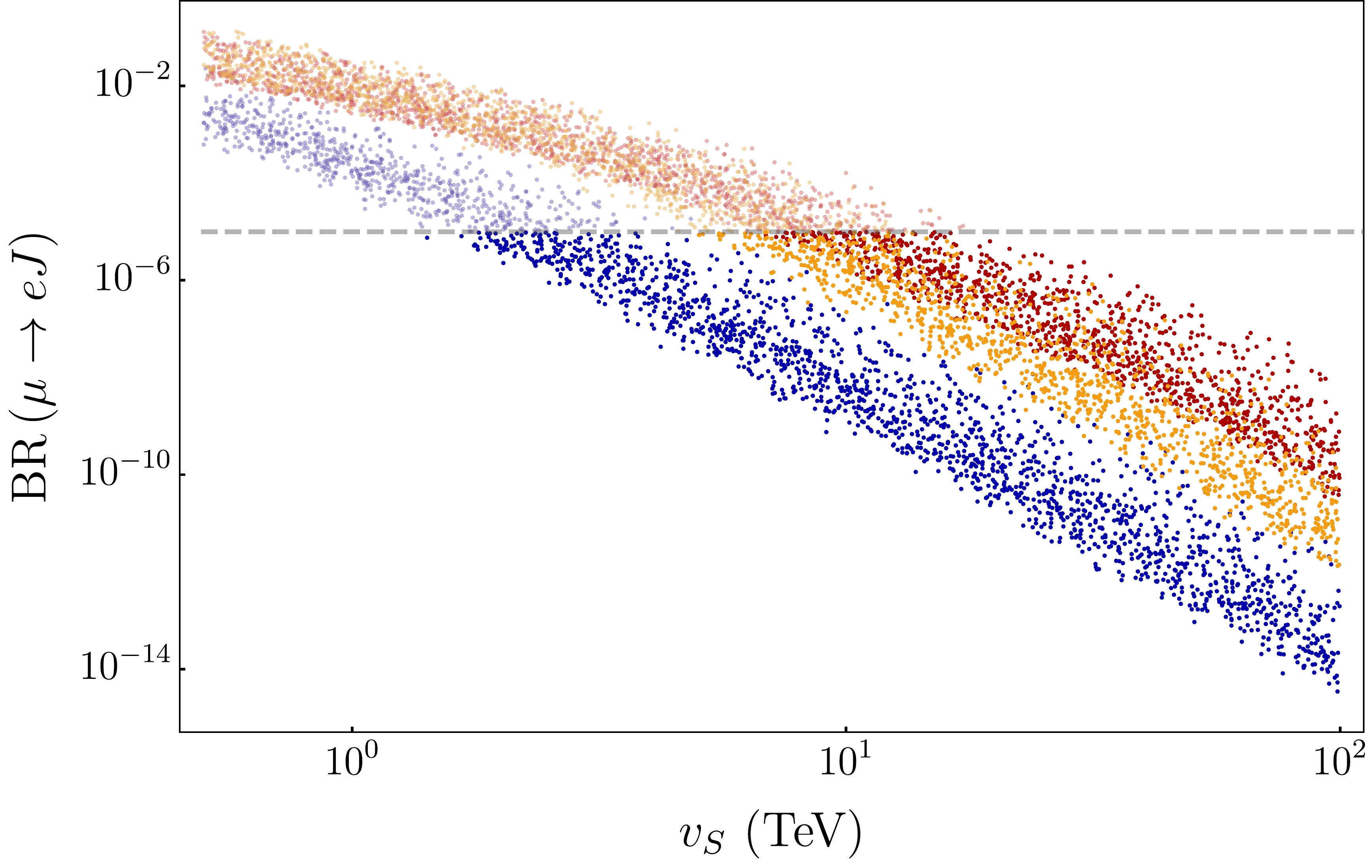}
  \caption{BR($\mu \to e \, J$) as a function of $v_S$ for three
    scenarios: $M_{N_n} = 0.5$ TeV, $m_\eta^+ \in [200,300]$ GeV
    (blue), $M_{N_n} = 5$ TeV, $m_\eta^+ \in [200,300]$ GeV (red) and
    $M_{N_n} = 0.5$ TeV, $m_\eta^+ \in [2,3]$ TeV (orange), with
    $\left(M_N\right)_{mn} = M_{N_n} \, \delta_{mn}$. See text for
    details.
    \label{fig:mueJ}
    }
\end{figure}

The majoron off-diagonal couplings to charged leptons induce flavor
violating decays, such as $\mu \to e \, J$. Using the general results
derived in~\cite{Escribano:2020wua} one can obtain predictions for the
model considered here. These are presented in Fig.~\ref{fig:mueJ},
which shows BR($\mu \to e \, J$) as a function of the lepton number
breaking scale $v_S$ for three scenarios: $M_{N_n} = 0.5$ TeV,
$m_\eta^+ \in [200,300]$ GeV (blue), $M_{N_n} = 5$ TeV, $m_\eta^+ \in
[200,300]$ GeV (red) and $M_{N_n} = 0.5$ TeV, $m_\eta^+ \in [2,3]$ TeV
(orange), with $\left(M_N\right)_{mn} = M_{N_n} \, \delta_{mn}$. All
points in this figure are compatible with neutrino oscillation
data. This has been achieved by using an adapted Casas-Ibarra
parametrization for the Yukawa matrix
$y$~\cite{Casas:2001sr,Toma:2013zsa} and taking neutrino oscillation
parameters in the $3 \sigma$ ranges determined by the global
fit~~\cite{deSalas:2020pgw}. The fermion singlet masses are set to
their numerical values by fixing $\kappa$ accordingly. Finally, the
scalar potential parameters have been randomly chosen, with the
exception of the effective $\lambda_5$ coupling that is fixed to
$-10^{-8}$. The horizontal dashed line displays the current
experimental limit BR($\mu \to e \, J$) $< 10^{-5}$ obtained by the
TWIST collaboration~\cite{TWIST:2014ymv}. This limit can be improved
by the Mu3e experiment by looking for a bump in the continuous Michel
spectrum. This strategy was recently shown to be able to rule out $\mu
\to e \, J$ branching ratios above $7.3 \times 10^{-8}$ at 90\%
C.L.~\cite{Perrevoortthesis}. Therefore, we conclude that our setup
leads to observable $\mu \to e \, J$ decays, which already rule out
part of the parameter space of the model and are detectable in the
near future. Qualitatively similar results are obtained for the
processes $\tau \to e \, J$ and $\tau \to \mu \, J$.

\subsubsection*{Collider signatures}

The light CP-even scalar $h$ is identified with the $125$ GeV state
discovered at the LHC, and thus we must guarantee that its properties
match those observed. In particular, its production cross-section and
decay modes must be (within the allowed experimental ranges) close to
those predicted for a pure SM Higgs boson. This can be generally
guaranteed if the mixing angle $\alpha$, defined in
Eq.~\eqref{eq:mixing}, is small. For instance, $h$ can decay invisibly
via $h \to J J$. The interaction Lagrangian of the CP-even scalar $h$
to a pair of majorons can be written as $\mathcal{L}_{h J J} =
\frac{1}{2} \, g_{h J J} \, h \, J^2$, with the dimensionful coupling
$g_{h J J}$ given by
\begin{equation} \label{eq:ghJJ}
  g_{h J J} = v_S \, \lambda_S \, \sin \alpha + \left( \lambda_3^S - \frac{v_H^2 \left| \lambda_{H S \Delta} \right|^2}{m_\Delta^2} \right) v_H \, \cos \alpha \, .
\end{equation}
Using the limit on the invisible Higgs branching ratio $\text{BR}(h
\to J J) < 0.11$ at $95\%$ C.L.~\cite{ATLAS:2020kdi}, assuming that
the total Higgs decay width is given by $\Gamma_h \approx
\Gamma_h^{\rm SM} = 4.1$
MeV~\cite{LHCHiggsCrossSectionWorkingGroup:2016ypw} and taking into
account that $\Gamma(h \to J J) = g_{h J J}^2/(32 \, \pi \, m_h)$, one
finds $g_{h J J} < 2.4$ GeV at $95\%$ C.L.. This constraint can be
easily satisfied by choosing $\lambda_3^S \lesssim 10^{-2}$. Finally,
the heavy CP-even scalar $\Phi$ can also be searched for at
colliders. However, for $\alpha \ll 1$ this state is mostly singlet
and has very suppressed production cross-sections at the LHC.

\subsubsection*{Dark matter}

The usual $\mathbb{Z}_2$ parity of the Scotogenic model is obtained in
our model as a remnant after lepton number breaking. As a consequence
of this, the lightest $\mathbb{Z}_2$-odd state is completely stable
and can be a good DM candidate. Two scenarios emerge: (i) fermion DM,
with the lighest singlet $N_1$ as DM candidate, and (ii) scalar DM,
with either $\eta_R$ or $\eta_I$ (depending on the sign of the
effective $\lambda_5$ coupling) as DM candidate. This is completely
equivalent to the minimal Scotogenic model. However, the new scalar
states at low energies can alter the DM phenomenology
substantially. For instance, in the case of fermion DM, more
constrained due to the strong bounds from lepton flavor violating
observables, see for instance~\cite{Vicente:2014wga}, it has been
shown that the annihilation channels $N_1 \, N_1 \, \to \, \text{SM}
\, \text{SM}$ and $N_1 \, N_1 \, \to \, J \, J$ may open up new viable
regions in the parameter space of the
model~\cite{Bonilla:2019ipe}. These s-channel processes, mediated by
the CP-even scalars of the model ($h$ and $\Phi$), have a strong
impact on the DM relic density, reducing the tuning normally required
in the original Scotogenic model with fermion DM.

\section{Summary and discussion}
\label{sec:conclusions}

An ultraviolet completion for the Scotogenic model has been proposed
in this letter. Our high-energy scenario contains additional degrees
of freedom which, after being integrated out, give rise to the
well-known low-energy Lagrangian of the Scotogenic model. In
particular, they induce a naturally small $\lambda_5$ coupling,
suppressed by two powers of the high scale $m_\Delta$. Our
construction also generates the dark $\mathbb{Z}_2$ parity of the
Scotogenic model, which emerges as a remnant symmetry at low
energies. In summary, our ultraviolet model addresses some of the
theoretical drawbacks of the original Scotogenic model.

In addition to the usual Scotogenic states, our setup predicts two
additional particles at low energies: a massive scalar and a massless
Goldstone boson, the majoron $J$. We have shown that they have a
remarkable impact on the phenomenology of the model. New processes in
flavor physics, such as $\mu \to e \, J$, are available and
detectable in the near future. The dark matter phenomenology is also
affected due to novel production mechanisms in the early
Universe. Finally, we also expect new signatures in colliders.

While we have concentrated on a model with a global $\rm U(1)_L$
symmetry, it is also interesting to consider a version of our setup in
which the $\mathbb{Z}_2$ parity has a gauge origin. In this case, the
majoron would be replaced by a massive $Z^\prime$ boson. Furthermore,
our ultraviolet completion is by no means unique and other models with
similar low-energy limit exist, perhaps with different
phenomenological predictions. We leave these possibilities for future
work.

\section*{Acknowledgements}

Work supported by the Spanish grants FPA2017-85216-P
(MINECO/AEI/FEDER, UE) and SEJI/2018/033 (Generalitat Valenciana). The
work of PE is supported by the FPI grant PRE2018-084599. AV
acknowledges financial support from MINECO through the Ramón y Cajal
contract RYC2018-025795-I.

%\appendix

%\input{tex/app}

\bibliographystyle{utphys}
\bibliography{refs}

\end{document}